\documentclass[twocolumn,superscriptaddress,aps,showpacs]{revtex4}
\usepackage{epsfig}
\usepackage{graphicx}
\usepackage{dcolumn}
\usepackage{bm}

\newcommand{\be}{\begin{equation}}
\newcommand{\ee}{\end{equation}}
\newcommand{\bea}{\begin{eqnarray}}
\newcommand{\eea}{\end{eqnarray}}
\newcommand{\om}{\omega}

\newcommand{\ii}{\left|i \right>}
\newcommand{\ff}{\left|f \right>}
\newcommand{\s}{\displaystyle{\not}}

\begin{document}

\preprint{APS/123-QED}

\title{Design of an electron microscope phase plate using a focused continuous-wave laser}

\author{H. M\"uller}
\address{Physics Department, University of California, Berkeley, CA
94720} \email{hm@berkeley.edu}
\author{Jian Jin}
\address{Engineering Division, Lawrence Berkeley National Laboratory,
University of California, Berkeley, CA 94720}
\author{R. Danev}
\address{Division of
Nano-Structure Physiology, Okazaki Institute for Integrative
Bioscience, 5-1 Higashiyama, Myodaiji, Okazaki 444-8787, Japan}
\author{J. Spence}
\address{Physics Department, Arizona State University, Tempe, AZ
85287-1504}
\author{H. Padmore}
\address{Advanced Light Source, Lawrence Berkeley National Laboratory
University of California, Berkeley, CA 94720}
\author{R.M. Glaeser}
\address{Life Sciences Division, Lawrence Berkeley National
Laboratory, University of California, Berkeley, CA 94720}


\begin{abstract}
We propose a Zernike phase contrast electron microscope that uses
an intense laser focus to convert a phase image into a visible
image. We present the relativistic quantum theory of the phase
shift caused by the laser-electron-interaction, study resonant
cavities for enhancing the laser intensity, and discuss
applications in biology, soft materials science, and atomic and
molecular physics.
\end{abstract}

\date{\today}
\maketitle

\section{Introduction}
The resolution of modern transmission electron microscopes (TEMs)
\cite{Books} can be more than 10,000 times the one of light
microscopes. Yet, their imaging performance for thin soft matter
such as biological specimens is limited: These specimens are
weakly scattering phase objects and a perfect image of them shows
almost no contrast. The image is imprinted, however, on the phase
of the electron's wave function. In optical microscopes, such a
phase image can be made visible by Zernike phase contrast
microscopy \cite{Zernicke}: the part of the light beam that has
not been diffracted by the specimen is phase shifted by $\pi/2$ by
a quarter-wave plate, thus converting phase modulation into
visible amplitude modulation.

Quarter-wave plates for TEMs have been suggested early
\cite{Boersch1947} but could not be fabricated until recently.
Examples are thin carbon film phase plates
\cite{Hosokawa2005,NagayamaDanev2008}, electrostatic (einzel lens)
designs \cite{Schroeder,Cambie2007}, or thin bar magnets
\cite{NagayamaDanev2009}. Although the most recent implementations
are greatly improved over earlier versions
\cite{DanevNagayama2001}, they are still prone to becoming
electrostatically charged when they are hit by an electron beam.
This causes distortion of the image. While it ought to be possible
to overcome this limitation by using suitable materials and by
keeping the surface of devices extremely clean, such solutions are
not necessarily well suited for routine TEM applications.
Moreover, microfabricated phase plates induce electron loss,
leading to reduced performance.

Attempts have also been made to use bi-prism electron holography to record the phase of the electron beam, but this approach has been largely abandoned because subtle charging effects in the specimen itself result in degraded temporal coherence in an off-axis hologram, whereas they have little effect on the in-line holography that is provided by Zernike phase-contrast imaging.

In view of these challenges, the
phase shift that occurs when an electron passes through a beam of
light \cite{Volkov} is here envisioned as an alternative way to
achieve phase contrast.
To provide the high required laser intensities, a pulsed laser can
be used together with a synchronized pulsed TEM \cite{llnl}.
However, using a pulsed-source TEM introduces new constraints for
routine TEM applications, e.g., lower averaged electron flux and
thus longer data collection periods, as the electron density
during a pulse cannot be higher than in cw operation to avoid
space charge effects, and potentially greater complexity.

In this Article, we propose designs for a continuous wave (cw)
version of an optical-beam quarter-wave plate that we believe will
be suitable for general use in a TEM. We will derive the phase
shift of an electron wave and discuss implementations that reach
the required small spot size and high intensity using focussing
optics and optical cavities. Finally, we will discuss applications
for biological specimens and other soft-matter specimens and give
an outlook on other potential applications.

\section{Origin of the phase shift}

The phase shift of the electron wave function caused by the laser
can be explained by the ponderomotive force, a repulsive force
felt by electrons entering the electric field of a laser. It is
given by \cite{BarwickBatelaan} $\phi=\hbar \alpha \lambda\rho
\delta t/m,$ where $\hbar$ is the reduced Planck constant,
$\alpha$ the fine structure constant, $\lambda$ the laser
wavelength, $\rho$ the photon density, $m$ the electron mass, and
$\delta t$ the time interval. This is a nonrelativistic result. In
the TEM, however, the electron energy is comparable to $mc^2$ and
thus relativistic corrections may be significant.


In order to obtain a relativistically correct result, we now
derive the phase shift using the theory of quantum electrodynamics
\cite{MandlShaw}. The electrons, long before the interaction with
the laser, are described by a quantum state $\ii$. A scattering
matrix $S=1+S^{(1)}+S^{(2)}+\ldots$ describes the evolution of
$\ii$ into the final state $\ff=S\ii$, which we define long after
the interaction. The first order $S^{(1)}$ describes emission or
absorption of a single photon by an electron. This violates energy
and momentum conservation, so $S^{(1)}=0$. It is important for us
that this still holds if the laser beam is focussed: the electric
field of the focussed beam can be described in the momentum
representation by a superposition of plane waves of various
directions. Since $S^{(1)}=0$ for each of these plane waves,
regardless of their direction or polarization, it is also zero for
the focussed beam.

\begin{figure}
\centering \epsfig{file=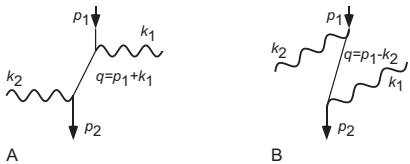,width=0.3\textwidth}
\caption{\label{Compton} Feynman diagrams for Compton scattering.
$p_1, q,$ and $p_2$ respectively denote the initial, intermediate,
and final momentum of the electron, $k_1$ and $k_2$ are the photon
momenta.}
\end{figure}

The leading nonzero effects are described by the second order,
$S^{(2)}$. Most of these take place whether or not the electrons
interact with the laser ({\em e.g.,} annihilation,
electron-electron scattering, self-energy, and vacuum polarization
\cite{MandlShaw}) and do not contribute to the phase shift of
interest. The remaining process (Fig. \ref{Compton}), called
Compton scattering, can transfer momentum and energy to the
electron. The scattering rate for such transfer happens is
$r=\sigma_T\rho v$, where $\sigma_T=[\alpha\hbar/(mc^2)]^2$ is the
Thomson cross section and $v$ the electron velocity. Since
electrons whose momentum has changed do not contribute to the TEM
image, Compton scattering with momentum transfer is a loss
mechanism.

The desired phase shift is caused by the degenerate case of
Compton scattering without momentum transfer. From the Feynman
diagrams in Fig. \ref{Compton}, the scattering matrix $S$ can be
calculated by following the Feynman rules, see Ref.
\cite{MandlShaw}, chapter 7.3. If we define $\phi\equiv
\left<f\right|S^{(2)}\ii$, the result can be expressed as (using
$\hbar=m=c=1$)
\bea
\phi&=& \frac{(2\pi)^4\delta^{(4)}(p_1+k_1-p_2-k_2)}{2V^2
\sqrt{E_{p_1}E_{p_2}\omega_{k_1}\omega_{k_2}}}(S_A+S_B), \nonumber
\\
S_A&=&-e^2\bar u(\vec p_2)\s \epsilon (\vec k_2)iS_F(q)\s\epsilon(\vec k_1)u(\vec p_1) a^\dag(\vec k_1)a(\vec k_2),\nonumber \\
S_B&=&-e^2\bar u(\vec p_2)\s \epsilon (\vec
k_1)iS_F(q)\s\epsilon(\vec k_2)u(\vec p_1) a^\dag(\vec k_2)a(\vec
k_1).
\eea
$S_A$ and $S_B$ denote terms arising from Fig. \ref{Compton}, A
and B. Here, $\delta^{(4)}$ is the four dimensional Dirac delta
function, $V$ the normalization volume, $E_p$ the energy of an
electron of 4-momentum $p$, $\omega_k$ the photon frequency, $S_F$
the Feynman propagator \cite{MandlShaw}, and $u$ and $\s\epsilon$
describe electron and photon polarization, respectively. Since no
momentum is transferred, $\vec k_1=\vec k_2$ and we identify
$\rho\equiv a^\dag(\vec k_1) a(\vec k_1)/V$ as the photon density
operator. Physically, this means that the emission of the outgoing
photon is stimulated by the laser, in contrast to the
non-degenerate case that is treated in most textbooks. We obtain
\cite{restframe} $\phi=-[ie^2\rho/(2E_p\om)] \bar u(\vec p) [(\s
k)/(p k)]  u(\vec p)\delta t$. In our TEM (Fig. \ref{lens}), the
electron beam travels along the $z$ axis and the laser
orthogonally along $y$. Furthermore, we choose the polarization of
the laser orthogonal to the direction of the electron beam. It
follows that $pk=-E_p\omega+\vec p\vec k=-E_p\omega$ and
\be\label{phiQM}
\phi =\frac{-e^2\rho\delta t}{2E_p^2\om^2} \bar u(\vec p)
\left[-\gamma^0 \omega+\gamma^2 k\right] u(\vec p)=\frac{\hbar
\alpha \rho\lambda\delta t}{\sqrt{m^2+p^2/c^2}}.
\ee
This agrees with the non-relativistic result if $p/c\ll m$.

If the laser is polarized parallel to the electron beam,
$pk=-E_p\omega+p_zk_z$ and $\phi =(1/2)\lambda\hbar \alpha \rho
\delta t/(p_z/c-\sqrt{m^2+p^2/c^2})$. For nonrelativistic
electrons, this is a factor of two lower than Eq. (\ref{phiQM}).

We remark that we have treated the laser beam as a plane wave. A
focussed beam could be described by a superposition of plane waves
in the momentum representation. We can treat each of them in the
above manner and calculate the combined effect. This will not lead
to strong modifications of our results, because (i) even for the
strongly focussed beams considered below, the major part of the
optical power propagates at wavevectors having relatively small
angles with the optical axis, and (ii) even for arbitrary
wavevectors, the phase shift caused is not zero or of opposite
sign, ruling out cancellations.

If $\phi\ll 1$, it clearly describes a phase shift:
$\ff=(1+i\phi)\ii\approx e^{i\phi}\ii$. To see that this still
holds for large $\phi$, we divide the time interval $\delta t$
into several small ones. Repeated application of the $S^{(2)}$,
once for each small interval, shows that $\ff=e^{i\phi}\ii$. This
process amounts to  summing an infinite series of Feynman
diagrams, in which the processes of  Fig. \ref{Compton} A, B
happen 1,2,3,\ldots times. This is valid as long as other
processes such as non-degenerate Compton effect remain negligible.
This is satisfied in our case: the fraction of electrons lost per
radian of phase shift is given by $r/\phi=\alpha
(v/c)(\lambda_C/\lambda)$. This is about $10^{-8}$ for
$\lambda=1\,\mu$m and thus negligible.

\section{Design of the laser phase plate}

\subsection{Near-spherical cavity}

\begin{figure}
\centering \epsfig{file=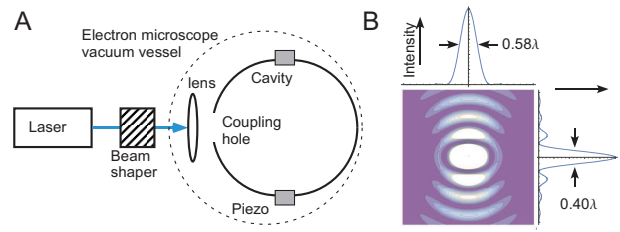,width=0.45\textwidth}
\caption{\label{TMn01mode} A: setup for a phase contrast TEM using
a near-spherical resonant cavity. The electron beam is propagating
orthogonal to the paper plane. B: Intensity of the TM$n$01 mode in
a spherical cavity. Shown is a 2-dimensional contour plot as well
as the intensities along the $x$ and $y$ axes through the focus,
respectively.}
\end{figure}

We now study how this phase shift can be applied for a cw phase
contrast TEM. The basic challenge is to reach a sufficient photon
density, which requires a large laser intensity. However, since
the phase shift is proportional to the laser wavelength $\lambda$,
we can use the longest wavelength at which we can still achieve a
focus of the required size. The smallest focus for a given
wavelength is achieved by the TM$n$01 mode in a spherical resonant
cavity \cite{Zhang2007} shown in Fig. \ref{TMn01mode} B. The
cavity provides both focussing as well as resonant enhancement of
the intensity: the laser has to supply only the power lost at the
cavity walls. The focus has a half-intensity radii of
$0.25\lambda$ and $0.20\lambda$ in $x$ and $y$-direction,
respectively; the largest intensity maxima other than the focus
are less then 10\% of the intensity at the focus and can be
neglected.

To determine the required laser power, we calculate the energy
flow $3 E_0^2 \pi/(4c\mu_0 k^2)$ towards the cavity walls, where
$E_0$ is the peak electric field at the focus, given the electric
field in the cavity \cite{Zhang2007}. The cavity walls reflect
most of this back, but $P=3(1-r)E_0^2\pi/(4\eta k^2)$, where $r$
is the reflectivity of the cavity walls is lost and has to be
replaced by the laser. Expressing the photon density by the
electric field $E_z^2= 2 \rho \hbar \omega c \mu_0$ (where the
factor of 2 accounts for the time averaging of the intensity) and
integrating along the z-axis $\int_{-\infty}^\infty
E_z^2(x=0,y=0,z)dz=\frac 35 E_0^2 \lambda$, we obtain
$\phi=\frac{4}{5(1-r)}\frac{ \lambda P\alpha} {\gamma c^2 m v}$,
where $\gamma=1/\sqrt{1-v^2/c^2}$, or
\be \label{numerical}
\frac{\delta}{\pi/2}\approx \gamma^{-1} \left(\frac{P}{{\rm
kW}}\right)\left(\frac{\lambda}{\mu{\rm m}} \right)
\left(\frac{v}{c}\right)^{-1}\mathcal N,
\ee
where $\mathcal N\approx 0.15/(1-r)$.

Such cavities can use metal coatings for the inner walls. For
gold, for example, $r\approx 0.98$ in the infrared. Achieving
optimal phase contrast with a cut-on spatial frequency of
1/(40\,nm) with a modified FEI Titan electron microscope that is
available at Berkeley requires a half-intensity radius of about 2
microns at an electron energy of 80\,keV or 1 micron at 300\,keV.
The longest wavelength at which a 2$\mu$m focus can be achieved is
about $\lambda\sim 10\,\mu$m. If we choose a CO$_2$ laser at
$10.6\,\mu$m and let $r=0.98$ and $v=c/2$, we need a power of
about 7.5\,W. A 1-micron focus can be achieved using $P=39$\,W at
$\lambda=2\,mu$m or $P=75$\,W at $1064\,$nm.

To help coupling the laser power into the TM$n$01 mode, distortion
of the sphere will lift the degeneracy of modes: the desired mode
can then be selected by tuning the laser frequency. Coupling the
laser beam into the cavity can be achieved, {\em e.g.}, via a hole
of radius $r_{\rm in}=R\sqrt{2(1-r)/3}$ in the cavity, where $R$
is the radius of the cavity. This radius is determined such that
the power lost at the walls balances the one delivered through the
hole when the electromagnetic field of the laser matches the
resonant field inside the sphere. The laser-cavity coupling
efficiency is then given by the overlap integral of the laser and
the cavity field. For a Gaussian laser beam, the optimum power
transfer is 50.4\% and occurs when the $1/e^2$ intensity radius
(`waist') of that beam equals the hole radius and a lens of
appropriate focal length is used (Fig. \ref{TMn01mode} A). The
overlap can be increased by transforming the Gaussian beam into a
uniform-intensity beam, which is possible with 84-90\% efficiency
\cite{Maitzen,Palima2008}. In order to maintain resonance, the
laser frequency can be stabilized to the cavity resonance (or vice
versa) using the Pound-Drever-Hall \cite{PDH} or similar methods.

\subsection{Plano-parabolic cavity}

A parabolic mirror with radius much larger than focal length
produces a focus with an electric field similar to the one in a
spherical cavity. Together with a planar partial mirror, a
plano-parabolic cavity can be realized, see Fig. \ref{lens}. This
is easier to manufacture, as no hollow spheres need to be
produced. Moreover, the partial mirror is planar or near-planar
and can thus have a low-loss dielectric coating. Finally, the
laser beam feeding the cavity is collimated, simplifying
mode-match. The phase shift is given approximately by the same
equation as for the sphere, but $r$ is replaced by an effective
reflectivity $r_{\rm eff}$, which is reduced because of losses
when the photons spill over the perimeter of the mirror. $r_{\rm
eff}=0.9$ might be possible, and the required laser power is now
35\,W. Such an arrangement could thus satisfy all requirements
using a CO$_2$ laser.

\subsection{Fabry-Perot cavity}

Power buildup by factors of 100,000 or more \cite{mirrors} are
possible with high-reflectivity dielectric mirrors. Unfortunately,
they can at present only be applied to relatively flat surfaces.
Cavities using such mirrors are called Fabry-Perot cavities; their
eigenmodes are Gaussian beams \cite{Siegman}. In order to keep
losses due to the mirror aperture below a part per million, we
choose the mirror radius $r_m \gtrsim (5/2) w(L/2)$, where
$w(L/2)$ is the $1/e^2$ intensity radius of the beam at the
mirrors. This results in a radius $w_0=(5/2) \lambda/(\pi{\rm
NA})$ of the focus, where ${\rm NA}=r_m/R$. The phase shift is
given by Eq. (\ref{numerical}) where the factor $\mathcal
N=0.030/(1-r)$.

At present, commercial dielectric mirrors reach NA=0.04. At
$\lambda=532\,$nm and a Finesse $\mathcal F\equiv \pi/(1-r)=
50,000$, we would thus need a laser power of 8\,W, which is
available from commercial lasers. However, to obtain an 1/2
intensity radius of 2\,$\mu$m, which corresponds to an $1/e^2$
intensity radius of $w_0=1.7\,\mu$m, a minimum NA=0.25 is required
at $\lambda=532$\,nm. If such a cavity could be built with a
Finesse of 7,500, it would reach the required phase shift with an
8\,W laser. Even then, however, the standing wave in the cavity
will have a large number of maxima and minima along the beam
direction, forming a phase grating that diffracts the electron
beam and is thus undesirable. Also, the half width of the
intensity in the $y$ direction is given by the Rayleigh range
$z_R\approx \pi \lambda/(2{\rm NA})^2$, and is even larger than
$w_0$. These disadvantages are alleviated at high NA$\approx 1$, which
makes the central maximum dominate.

\subsection{Designs without cavity}

\begin{figure}
\centering \epsfig{file=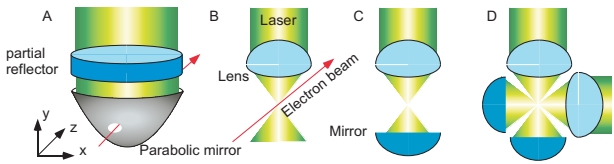,width=0.45\textwidth}
\caption{\label{lens} Alternative realizations of laser phase
plates. A: plano-parabolic cavity; B: high-NA lens; C: high-NA
lens with retroreflector; D: two focused, retroreflected beams.
The electron beam is propagating orthogonally to the page.}
\end{figure}

A simple way to achieve this
is a lens of large numerical aperture NA, see Fig. \ref{lens}, not
using a cavity. If the aperture of the lens is 3/2 times the waist
$w(f)$ at the lens (so that it transmits 99\% of the laser power)
the resulting radius of focus is $w_0\approx \lambda/(2$NA). The
phase shift is given by Eq. (\ref{numerical}) with $\mathcal
N=0.048 $NA. With a NA of 1, the maximum wavelength in this
implementation is about 3$\,\mu$m. Unfortunately, a power of about
4\,kW would be required even if NA=1. At $\lambda=10.6\,\mu$m with
NA=1, this is reduced to $P=1.2$\,kW is required, but this results
in $w_0=5\,\mu$m, about 3 times too large.

In Fig. \ref{lens} C, a standing wave is generated, which has a
half-intensity half-width of $\lambda/8$ in the beam direction at
the cost of a more complicated intensity profile that has several
minor peaks. If the focal point of the lens and the mirror
coincide, the electric field there is doubled and the intensity
quadrupled, resulting in $\mathcal N=0.19$\,NA. To obtain low spot
size in two dimensions, two standing waves can be overlapped as
shown in Fig. \ref{lens}, D. If these two beams have polarizations
orthogonal to the electron beam, and thus orthogonal to each
other, their electric fields add geometrically and $\mathcal
N=0.19$NA. A similar implementation might use the focused spot as
in Fig. \ref{lens} B and use mirrors to turn the beam 270 degrees
and bring it in through a second lens, similar to Fig. \ref{lens}
D. We remark that at high NA, the theory of Gaussian beams becomes
inaccurate \cite{LiebMeixner}. Depending on the combination of
optics and polarization, a factor of safety of 1.2-2 should be
applied to the above estimates at NA=1. This is a further
advantage of the spherical or plano-parabolic cavity, where exact
analytical expressions have been used to calculate the fields.



\section{Applications}

At present, the spherical or plano-parabolic cavities
appear to be the most promising approaches to construct a cw laser
phase plate. Such a phase plate has a number of important
advantages over microfabricated ones: Since it does not use
mechanical structures inside or near the electron beam, it can be
completely free of the unwanted electrostatic charging that can
occur when electrons hit a solid-phase target, under conventional
vacuum conditions. Also, almost none of the electrons that pass
through the optical beam are lost, in contrast to thin film or
microfabricated phase plates. Moreover, laser phase plates are not
subject to the aging experienced by thin carbon-film phase plates
\cite{Danev2009}.

Development of such a reliable phase plate would significantly
expand the usefulness of TEMs in biology and in soft-materials
science.  In current practice, the only option for generating
contrast for thin organic materials like biological specimens is
to emulate a phase plate by intentionally using a rather large
amount of defocus, combined with spherical aberration
\cite{Lentzen2004}. Contrast transfer in a defocused TEM image is
quite poor, for example, for the low spatial frequencies that
carry most of the information about the size, shape, and location
of an object; it also suffers from contrast reversals (and
accompanying zeros) as well as from a damped envelope at higher
resolution, when the amount of defocus is increased enough to
improve the contrast transfer at low frequencies.  By contrast,
the laser phase plate would provide full image contrast at low and
high resolution at the same time.
Use of a quarter-wave phase plate in biology will make it possible
to image much smaller protein complexes than can currently be
visualized by defocus-based phase contrast, as has been
demonstrated with the use of a thin-film phase plate
\cite{Danev2009}. In most soft-matter applications, the density
differences are even smaller than they are for biological
materials that are embedded in vitreous ice, and thus it is hardly
possible to image polymer blends, etc. with defocus-based phase
contrast. In soft-matter applications, the use of a quarter-wave
phase plate will thus open up greater possibilities for imaging
unstained specimens under low-dose conditions.

\section{Outlook}

As an outlook, the intense 3 dimensional laser foci generated for
this work can also be used as very deep dipole traps for electrons
\cite{Moore}, atoms, and molecules. Trap depths could be in the
range of tens or even hundreds of Kelvin, allowing, for example,
to trap room-temperature atoms of almost any kind and localize
them to better than 0.5 microns. This would allow for spectroscopy
of exotic species. For example, spectroscopy of the $\sim$5-6 eV
transition in the Thorium 229 nucleus would allow for the
construction of clocks based on nuclear energy levels but has so
far been thwarted by the lack of a suitable atom trap. The dipole
trap proposed here could solve this problem and thus lead to
higher precision clocks (as the nucleus of an atom is less
sensitive to the environment than the electron shell), tests of
the time variability of fundamental constants of unprecedented
accuracy, or searches for an electron electric dipole moment
\cite{EDM}. Moreover, a phase contrast TEM could be used for
scattering-free quantum nondemolition imaging of atoms, ions, or
molecules.

\acknowledgments

We thank Eva Nogales for discussions and Mike Hohensee for help in preparing the manuscript. This research was supported by NIH grant GM083039, the David and Lucile Packard Foundation, and the Alfred P. Sloan foundation.






\end{document}